# The effect of hydroxyl on dye-sensitized solar cells assembled with TiO$_2$ nanorods

Lijian Meng(孟立建)[a,b], Tao Yang(杨涛)[c], Sining Yun(云斯宁)[a] and Can Li(李灿)[d]

a) Functional Materials Laboratory, School of Materials & Mineral Resources, Xi'an University of Architecture and Technology, China
b) Departamento de Física and Centre of Innovation in Engineering and Industrial Technology, Instituto Superior de Engenharia do Porto, Instituto Politécnico do Porto, Portugal.
c) Key laboratory of Biofuels, Qingdao Institute of Bioenergy and Bioprocess Technology, Chinese Academy of Sciences, China
d) State Key Laboratory of Catalysis, Dalian Institute of Chemical Physics, Chinese Academy of Sciences, China.

**Abstract**

TiO$_2$ nanorods have been prepared on ITO substrates by dc reactive magnetron sputtering technique. The hydroxyl groups have been introduced on the nanorods surface. The structure and the optical properties of these nanorods have been studied. The dye-sensitized solar cells (DSSCs) have been assembled using these TiO$_2$ nanorods as photoelectrode. And the effect of the hydroxyl groups on the properties of the photoelectric conversion of the DSSCs has been studied.

**Key words:** Hydroxyl groups, TiO$_2$ nanorod, DSSC, Sputtering

**PACS:** 68.37.-d, 81.15.Cd, 88.40.H-

---

[] Corresponding author. email: ljm@isep.ipp.pt

# 1. Introduction

Dye-sensitized solar cells (DSSCs) have attracted significant attention due to their special features, such as low cost and high light to electricity conversion efficiency. The cells generally are composed by a dye adsorbed nanoporous material, typically $TiO_2$, and an electrolyte solution as a hole transport layer containing a dissolved iodide ion/tri-iodide ion redox couples [1]. The dye molecules absorb light to generate excited electron-hole pairs. The electrons are then injected into the porous $TiO_2$ photoelectrode and propagate through it until they are collected and transferred to the external electric circuit. The electron transport, recombination and collection processes are three very important processes in DSSC and have been extensive studied [2-18]. In order to improve the conversion efficiency, the charge recombination possibility must be reduced. Therefore a high efficiency and fast charge transportation process is required. One dimension (1D) nanostructure, such as nanorod, nanotube and nanowire, shows a promising solution to improve the charge transportation process. Electron transport in 1D structure is expected to be several orders of magnitude faster than in random 3D nanostructure[12, 19, 20]. Many works have been done for 1D structure based DSSC and the conversation efficiency is approaching that for 3D nanoparticles based DSSC [21-27].

Traditionally, the photoelectrodes of DSSC are made by chemical method which needs a high temperature (450 ºC) treatment for the densification of the $TiO_2$ films. This high temperature treatment will cause the problem for producing the flexible cells as the polymeric substrates could not endure such a high temperature. Magnetron sputtering technique has been considered as industrial processes that are applicable to large scale deposition with high uniformity at a relatively low deposition temperature. In the beginning of this century, some works on DSSC using $TiO_2$ electrode prepared by sputtering technique have been reported by Goméz et.al. [28, 29]. Recently, some other groups have also reported the results on DSSC based on $TiO_2$ films prepared by sputtering technique [30-33]. So far, the energy conversion efficiency is still low for the DSSCs based on sputtered $TiO_2$ films as it cannot adsorb a large amount of dye molecules because of the lower specific surface area resulting from the compact structure which is a typical characteristic of the sputtered films. In our previous work, the $TiO_2$ nanorods have been made by dc reactive magnetron sputtering and the effects of the nanorods dimension, blocking layer and annealing temperature on the efficiency of DSSCs have been reported [34-38]. It is well known that the ability to adsorb photosensitive dyes can be improved by introducing the surface hydroxyl groups. In this work, the effect of the surface hydroxyl groups on the structure of the nanorods is discussed and the DSSCs were assembled using these $TiO_2$ nanorods as the electrode. The photovoltaic properties of DSSCs are studied.

## 2. Experimental section

The titanium oxide nanorods were deposited both on glass and commercial ITO substrates by dc reactive magnetron sputtering technique. The nanorods deposited on glass substrates were used for the simulations of the transmittance for obtaining the film thickness and the optical constants. The nanorods deposited on ITO substrates were used for dye-sensitized solar cells. The resistance of the ITO substrate is 30 -- 40 Ω per square. The titanium metal with a purity of 99.99% (Φ 60 x 3 mm, Grikin Advanced Materials Co. Ltd.) was used as the sputtering target. The vacuum chamber was pumped using a turbo molecular pump backed with a mechanic pump. Before the deposition, the chamber was pumped down to $1 \times 10^{-3}$ Pa, and then high purity Ar and $O_2$ gases were introduced into the chamber through the individual mass flow controller. The oxygen partial pressure and the total sputtering pressure in the chamber were kept at 0.3 Pa and 1.5 Pa, respectively for all deposition processes. The target-substrate distance was kept at 60 mm. No extra heating and biasing has been applied for the substrate during all the deposition processes. The sputtering current and the cathode potential were kept at 0.5 A and 400 V, respectively for all the depositions. The deposition time was 6 hours. The hydroxyl group were introduced by passing the oxygen gas through the water before it was introduced into the chamber. The water was kept in the room temperature before it was introduced into the chamber. We did not measure the temperature of the water after it was introduced into the chamber. It is suggested that the temperature would be increased due to the effect of the plasma in the chamber. The sample prepared by this method was designated as sample prepared with water and the sample prepared by normal condition was designated as sample prepared without water.

The transmittance of the films was measured using a Jasco V-550 UV-Vis spectrophotometer. The film thickness and the optical constants have been calculated by fitting the transmittance using Scout software. The XRD measurements have been done using Rigaku miniflex goniometer (30 kV, 15 mA). The surface morphologies were studied using field emission scanning electron microscope (FE-SEM). In order to get the clear images, the low vacuum model has been used. X-ray Photoelectron Spectroscopy (XPS) was recorded on a Thermo Escalab 250 equipped with a monochromatic Al Kα X-ray source. The spectra were analyzed using CasaXPS (Casa Software, Ltd.). A standard

Shirley baseline without any offset was used for background correction. The C 1s spectrum for adventitious carbon (284.8 eV) was used for charge correction. The deposited $TiO_2$ films were sensitized with N719 dye by soaking the films in an ethanolic solution of the N719 dye (0.5 mM of $(Ru(II)L_2(NCS)_2:2TBA$, where L = 2,2'-bipyridyl-4,4'-dicarboxylic acid) for 24 hours at room temperature. The counter-electrode is sputtered Pt on the FTO glass and the electrolyte is composed of 0.1 M $I_2$, 0.1 M LiI, 0.6 M 1-hexyl-3-methylimidazolium iodide, and 0.5 M 4-tert-butylpyridine in 3-methoxypropionitrile. The photocurrent-voltage measurements were carried out with a princeton 2273 applied research electrochemical system, a 500 W xenon lamp under AM 1.5G (100mWcm$^{-2}$) illumination and a water filter. The light intensity was adjusted to 100 mW/cm$^2$. Cells were tested using a metal mask with an active area of 0.15 cm$^2$.

## 3. Results and Discussion

In order to see the formation of the hydroxyl group in the sample surface, the detailed XPS spectra of O 1s and Ti 2p for $TiO_2$ nanorods prepared with and without water have been measured as shown in Fig. 1. The O 1s peak can be deconvoluted by three peaks located at 529.8 eV, 531.6 eV and 533.2 eV respectively. The strongest peak located at 529.8 eV can be attributed to O-Ti bonding ($O^{2-}$), while the other two peaks located at a higher energy side (531.6 eV and 533.2 eV) can be attributed to the hydroxyl groups ($OH^-$) and hydrate and/or adsorbed water ($H_2O$). It can be seen that the intensity of the peaks related with OH groups is much higher for the sample prepared with water than the sample prepared without water. It can be suggested that the hydroxyl groups have been formed by introducing the water during the sputtering process and result in a high OH group peak intensity. For Ti 2p, two peaks located at 458.2 eV and 463.8 eV are detected, which correspond to Ti $2p_{3/2}$ and $2p_{1/2}$, respectively. The samples prepared with and without water show a similar Ti 2p spectra. It indicates that the introducing of the hydroxyl groups does not affect Ti 2p spectra.

SEM images of the $TiO_2$ nanorods prepared with and without water are shown in Fig. 2. It can be seen that the samples prepared with and without water have a similar nanorod structure. The increase of the presence of surface hydroxyl group does not affect the $TiO_2$ nanorod structure.

Figure 3 shows the XRD patterns of $TiO_2$ nanorods prepared with and without water. All the peaks in the XRD patterns can be indexed as anatase phase of $TiO_2$ and the diffraction data were in good agreement with PDF card 21–1272. No other phase of $TiO_2$ has been observed. It can be seen that the $TiO_2$ nanorods prepared with and without water have a preferred orientation along the [110] direction. However, the $TiO_2$ nanorods prepared with water have a stronger (220) diffraction peak intensity than that prepared without water. The intensity of the (101) diffraction peak is similar for two samples. The (101) and the (220) peak intensities have been calculated by fitting the XRD patterns and the ratio of the I(220)/I(101) for the nanorods prepared with and without water has been obtained. The ratio is 61 and 24 for nanorods prepared with and without water. From the Fig. 3 it can be seen that the (101) peak intensity does not have a very clear change for samples prepared with and without water. It means that the preferred orientation along the [110] direction is enhanced for the $TiO_2$ nanorods prepared with water.

The average surface energies of the crystal planes of the anatase $TiO_2$ are related to the percentage of the 5-foldcoordinated titanium atoms on the specific planes and are 1.09 $J/m^2$, 0.90 $J/m^2$, 0.53 $J/m^2$ and 0.44 $J/m^2$, respectively, for the [110], [001], [100] and [101] crystal planes [39]. Usually, the [101] planes dominate anatase $TiO_2$ single crystal, which are thermodynamically stable due to a low surface energy. However, Both the surface energy and the strain energy of grains formed in the films will affect the development of the texture for the polycrystalline films. The effects of strain energy minimization are qualitatively similar to those of surface and interface energy minimization in that normal grain growth can not occur until the subpopulation of grains favored by strain energy minimization has consumed all grains with other orientations[40]. The competition between surface energy and strain energy during film growth might contribute to the changes in texture of the grains as observed in Fig. 3. For sufficiently thin films, surface and interface energy minimizing textures are favored but for the thicker films with higher elastic strains, strain energy minimizing textures are formed[40]. It means that the (110) texture is dominated by strain energy minimization and the (101) texture is dominated by surface energy minimization in the growth process. By introducing water during sputtering process, the mobility of the adatoms in the substrate might be low and results in strain energy minimizing textures favorable during grain growth and a high (220) diffraction peak intensity.

Figure 4 shows the specular transmittance spectra of the $TiO_2$ nanorods prepared with and without water. The transmittance of the nanorods prepared with water is slightly higher than the nanorods prepared without water. By fitting the transmittance, the dispersions of the refractive index can be extracted. The results are shown in Fig.5. The nanorods prepared without water have a higher refractive index than that prepared with water. The refractive index is related with the packing density of the sample. The hydroxyl groups in the sample surface may result in a decrease of the packing density and a low refractive index.

Figure 6 shows the absorption spectra of the $TiO_2$ nanorods prepared with and without water. It can be seen that the dye absorption (around 500 nm wavelength) is much higher for $TiO_2$ nanorods prepared with water than that prepared without water. Although the specific surface area is not measured, the morphology and the dimension of the nanorods are very similar for the samples prepared with and without water as shown in Fig. 2. It means that the specific surface area does not change with the introduction of the hydroxyl. Therefore, the increase of the dye absorption results from the formation of the hydroxyl groups in the surface of the nanorods. The J-V curves of DSSCs using $TiO_2$ nanorods prepared with and without water as photoelectrode are plotted in Figure 7 and the operation parameters are summarized in Table 1. The photoelectric conversion efficiency was calculated using the equation:

$$\eta = \frac{J_{sc}V_{oc}FF}{P_{in}} * 100$$

where η is the conversion efficiency, $J_{sc}$ is short circuit current density which depends on the charge injection and transportation, $V_{oc}$ is open circuit voltage which is most likely related with the difference between the Fermi level of semiconductor electrode and redox potential in the electrolyte, FF is fill factor which is related to functioning of the $TiO_2$/electrolyte interface, the higher the recombination of conduction band electrons with the electrolyte, the lower will be FF[12], and $P_{in}$ is incident light energy. It can be seen clearly that the solar cell assembled with $TiO_2$ nanorods prepared with water has a higher photocurrent density comparing to that prepared without water. The presence of the hydroxyl groups increases the absorption of the dye molecular and result in a high photocurrent. However, a negative effect is also be introduced by introducing the hydroxyl groups on the nanorod surface. As it can be seen from the Table 1 that the fill factor FF is lower for solar cell assembled with nanorods prepared with water, which may reduce the conversion efficiency. It is no yet very clear why the introduction of the hydroxyl reduces the fill factor. It is suggested that the introduction of the hydroxyl will modify the parasitic resistances of the cell and result in a low fill factor. The conversion efficiency is dominated by photocurrent. By introducing the hydroxyl groups on the nanorods surface, the efficiency is increased from 3.1% to 3.8%.

## 4. Conclusions

$TiO_2$ nanorods were prepared by dc reactive magnetron sputtering. The hydroxyl groups on the nanorod surface were introduced by passing the oxygen reactive gas through water. The preferred orientation along the [110] direction has been enhanced and the dye absorption has been improved by the hydroxyl groups. The DSSCs assembled using $TiO_2$ nanorods with hydroxyl groups show a better conversion efficiency than those using $TiO_2$ nanorods without hydroxyl groups.

# References

[1] O'Regan B and Gratzel M 1991 Nature **353** 737

[2] Wang H X and Peter L M 2012 J Phys Chem C **116** 10468

[3] Wang H X, Liu M N, Zhang M, Wang P, Miura H, Cheng Y and Bell J 2011 Phys Chem Chem Phys 13 17359

[4] Burke A, Ito S, Snaith H, Bach U, Kwiatkowski J and Gratzel M 2008 Nano Lett 8 977

[5] Gu Z Y, Gao X D, Li X M, Jiang Z W and Huang Y D 2014 J Alloy Compd 590 33

[6] Zhao X G, Jin E M and Gu H B 2013 Appl Surf Sci 287 8

[7] Zhang X B, Tian H M, Wang X Y, Xue G G, Tian Z P, Zhang J Y, Yuan S K, Yu T and Zou Z G 2013 Mater Lett 100 51

[8] Zhang X H, Ogawa J, Sunahara K, Cui Y, Uemura Y, Miyasaka T, Furube A, Koumura N, Hara K and Mori S 2013 J Phys Chem C 117 2024

[9] Zhang X B, Tian H M, Wang X Y, Xue G G, Tian Z P, Zhang J Y, Yuan S K, Yu T and Zou Z G 2013 J Alloy Compd 578 309

[10] Kuo Y Y, Lin J G and Chien C H 2012 J Electrochem Soc 159 K46

[11] Kwon Y S, Song I Y, Lim J, Park S H, Siva A, Park Y C, Jang H M and Park T 2012 Rsc Adv 2 3467

[12] Thavasi V, Renugopalakrishnan V, Jose R and Ramakrishna S 2009 Mat Sci Eng R 63 81

[13] Park J T, Chi W S, Jeon H and Kim J H 2014 Nanoscale 6 2718

[14] Wu M X, Lin X, Wang Y D, Wang L, Guo W, Qu D D, Peng X J, Hagfeldt A, Gratzel M and Ma T L 2012 134 3419

[15] Wu M X, Lin X, Wang T H, Qiu J S and Ma T L 2011 Energ Environ Sci 4 2308

[16] Liang D W, Tang Q W, Chu L, Li Q H, He B L, Cai H Y and Wang M 2013 Rsc Adv 3 25190

[17] He B L, Tang Q W, Luo J H, Li Q H, Chen X X and Cai H Y 2014 J Power Sources 256 170

[18] Gao Y F, Nagai M, Seo W S and Koumoto K 2007 J Am Ceram Soc 90 831

[19] Zhu K, Neale N R, Miedaner A and Frank A J 2007 Nano Lett 7 69

[20] Law M, Greene L E, Johnson J C, Saykally R and Yang P D 2005 Nat Mater 4 455

[21] Adachi M, Murata Y, Takao J, Jiu J, Sakamoto M and Wang F 2004 J Am Chem Soc, 126 14943

[22] Jiu J, Isoda S, Wang F and Adachi M 2006 J Phys Chem B 110 2087.

[23] Shao F, Sun J, Gao L, Chen J Z and Yang S W 2014 Rsc Adv 4 7805

[24] Sabba D, Agarwala S, Pramana S S and Mhaisalkar S 2014 Nanoscale Res Lett, 9 14

[25] Ren J B, Que W X, Yin X T, He Y C and Javed H M A 2014 Rsc Adv 4 7454.

[26] Ngamsinlapasathian S, Sakulkhaemaruethai S, Pavasupree S, Kitiyanan A, Sreethawong T, Suzuki Y and Yoshikawa S 2004 J Photoch Photobio A 164 145

[27] Lee B H, Song M Y, Jang S Y, Jo S M, Kwak S Y and Kim D Y 2009 J Phys Chem C 113 21453

[28] Gomez M M, Lu J, Olsson E, Hagfeldt A and Granqvist C G 2000 Sol Energ Mat Sol C 64 385

[29] Gomez M M, Lu J, Solis J L, Olsson E, Hagfeldt A and Granqvist C G 2000 J Phys Chem B, 104 8712

[30] Hossain M F, Biswas S, Takahashi T, Kubota Y and Fujishima A 2008 Thin Solid Films 516 7149

[31] Waita S M, Aduda B O, Mwabora J M, Granqvist C G, Lindquist S E, Niklasson G A, Hafeldt A and Boschloo G 2007 J Electroanal Chem 605 151

[32] Kang S H, Kang M S, Kim H S, Kim J Y, Chung Y H, Smyri W H and Sung Y E 2008 J Power Sources 184 331

[33] Sung Y M and Kim H J 2007 Thin Solid Films 515 4996

[34] Meng L J, Ma A, Ying P, Feng Z and Li C 2011 J Nanosci Nanotechnol 11 929

[35] Meng L J and Li C 2011 Nanosci Nanotech Let 3 181

[36] Meng L J, Li C and dos Santos M P 2011 J Inorg Organomet P 21 770

[37] Meng L J, Li C and dos Santos M P 2013 J Inorg Organomet P 23 787

[38] Meng L J, Ren T and Li C 2010 Appl Surf Sci 256 3676

[39] Lazzeri M, Vittadini A and Selloni A 2002 Phys Rev B 65 119901

[40] Thompson C V 2000 Annu Rev Mater Sci 30 159

Table 1 Performance comparison of the DSSCs assembled with TiO$_2$ nanorods prepared with and without water.

|  | $J_{sc}$ (mA/cm$^2$) | $V_{oc}$ (V) | FF | η |
|---|---|---|---|---|
| Without water | 8.4 | 0.64 | 0.58 | 3.1% |
| With water | 12.1 | 0.63 | 0.50 | 3.8% |

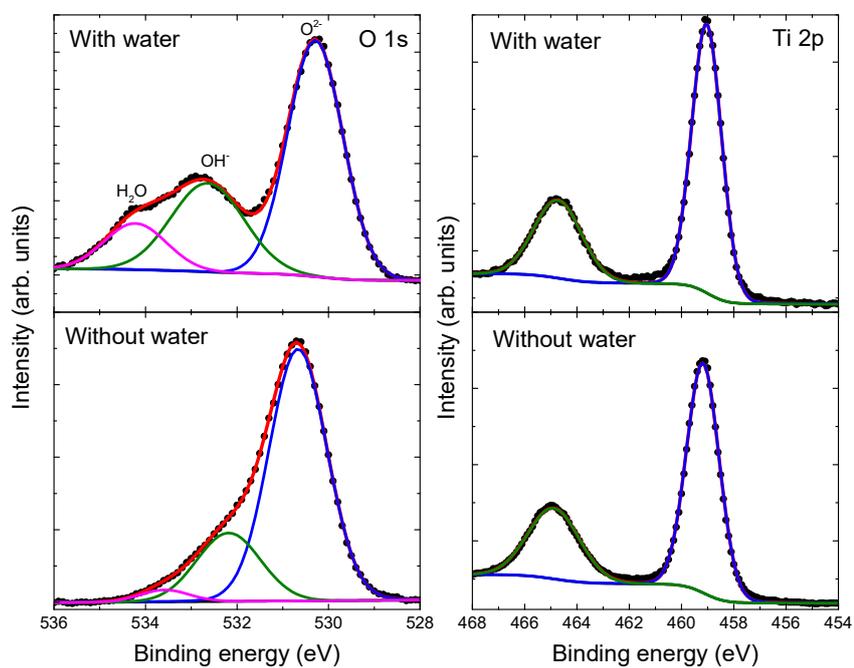

Figure 1. High-resolution XPS spectra for the O 1s and Ti 2p peaks and respective fitting of the $TiO_2$ nanorods prepared with and without water.

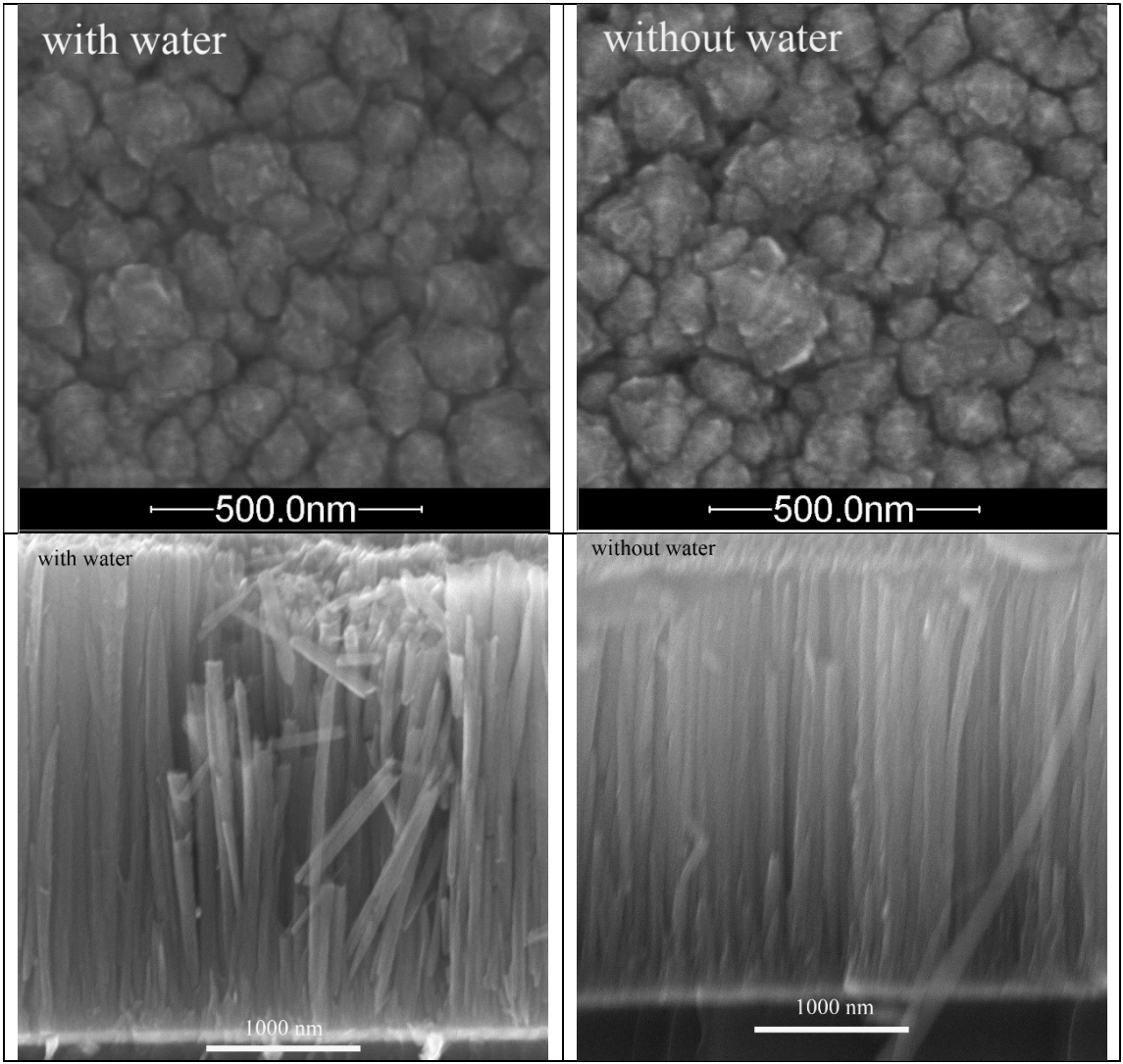

Figure 2, The SEM images of TiO$_2$ nanorods prepared with and without water.

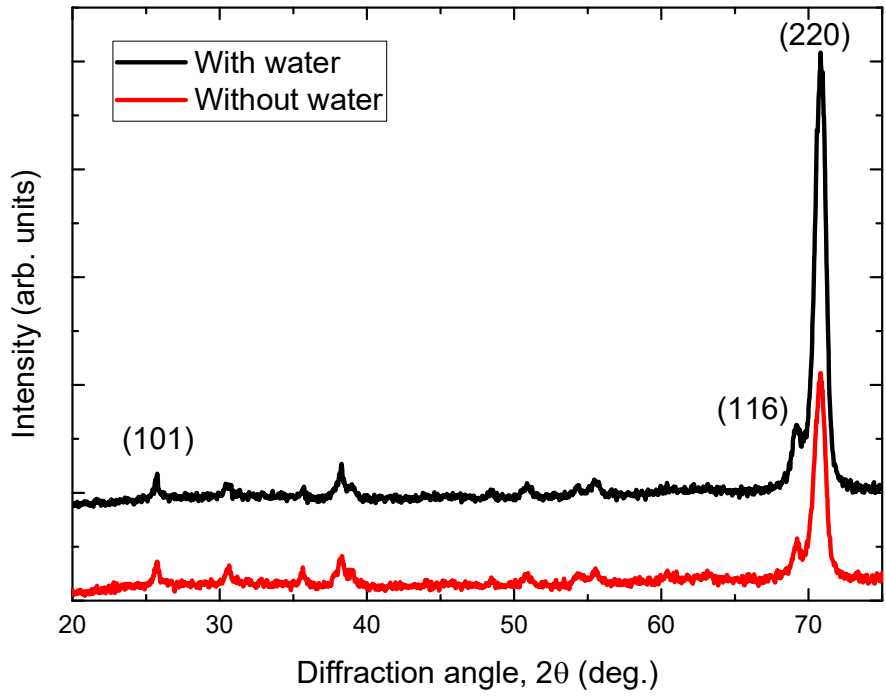

Figure 3. The X-ray diffraction patterns of TiO$_2$ nanorods deposited onto ITO substrates with and without water.

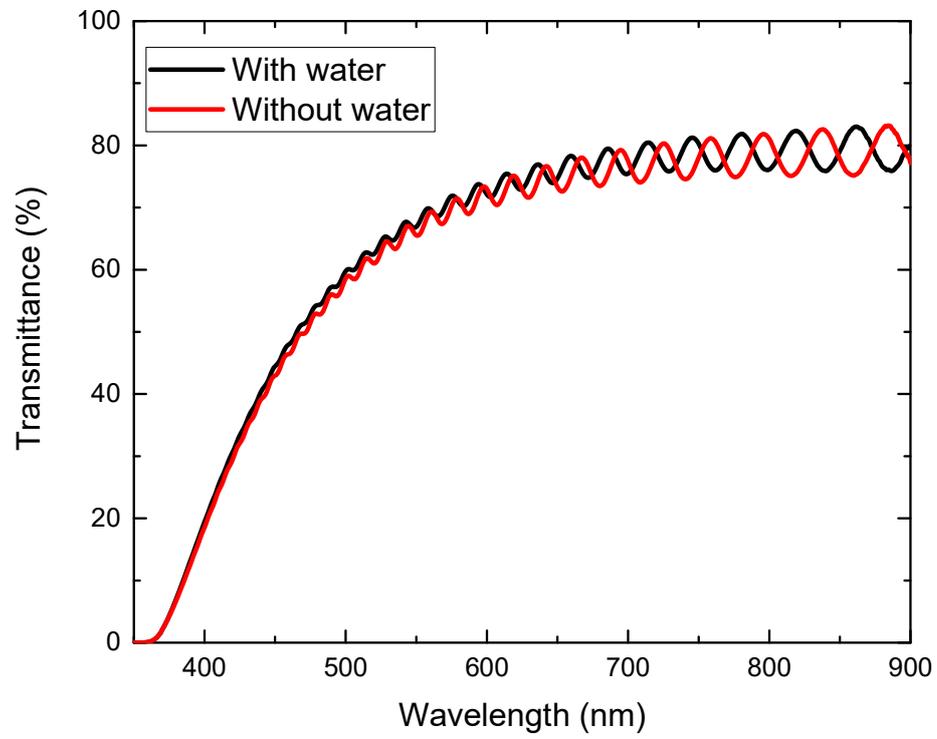

Figure 4. The transmittance of the $TiO_2$ nanorods deposited on glass substrate with and without water.

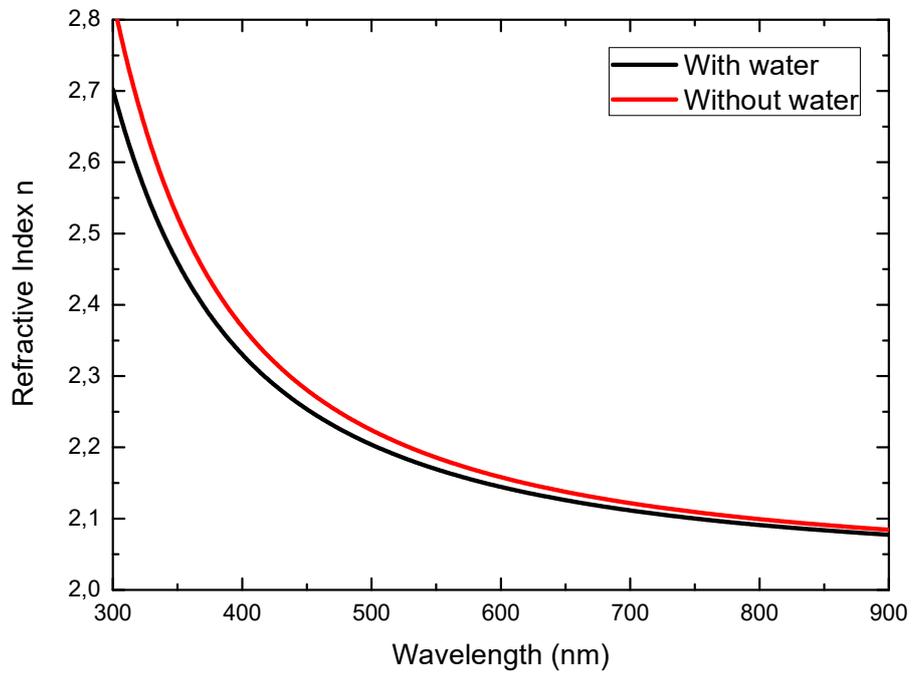

Figure 5. The refractive index of the TiO$_2$ nanorods prepared with and without water.

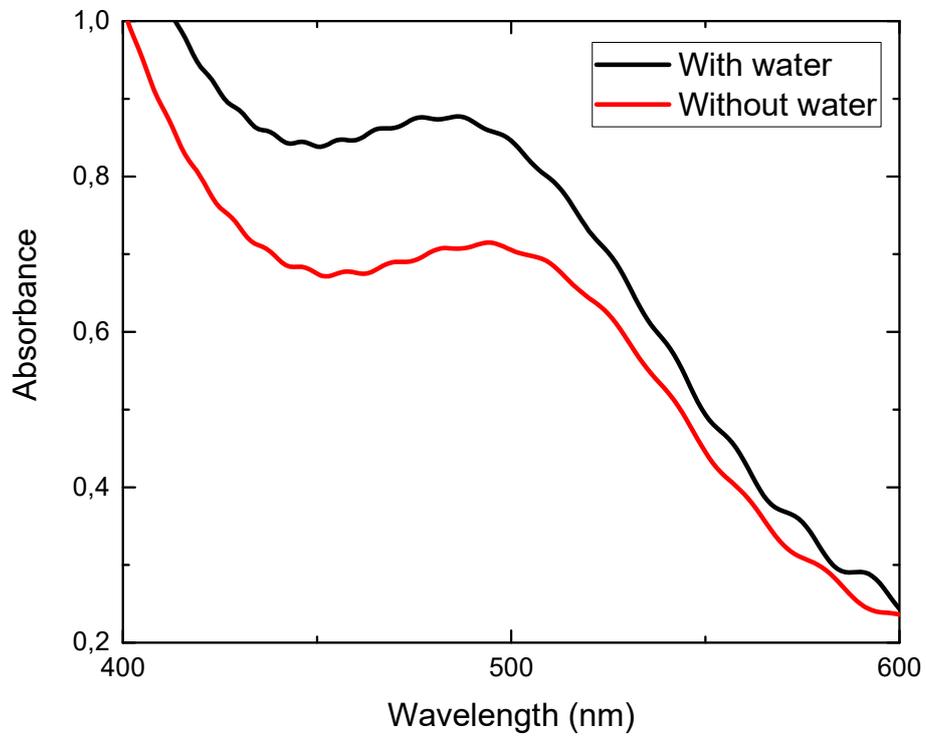

Figure 6. Absorption spectra of TiO$_2$ nanorods prepared with and without water after dye sensitized.

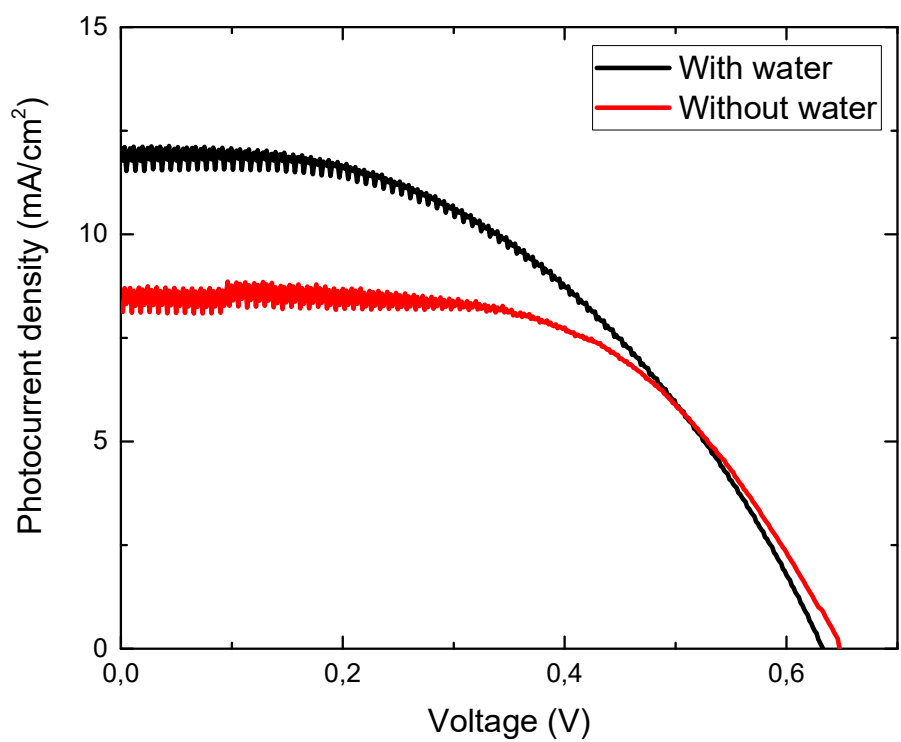

Figure 7. Current-potential characteristics of the TiO$_2$ nanorods prepared with and without water.